\documentclass[11pt]{article}
\usepackage[textwidth=460pt,textheight=660pt]{geometry}
\usepackage[usenames,dvipsnames]{color}
\usepackage{fancyvrb}
\usepackage{relsize}
\usepackage{amsmath}
\usepackage{amssymb}
\usepackage{graphicx}
\usepackage[urlcolor=blue,citecolor=blue,colorlinks]{hyperref}
\usepackage{fleqn}
\usepackage{xspace}
\usepackage[bottom]{footmisc}
\usepackage[numbers,sort&compress]{natbib}
\usepackage{tableaux}

\fvset{frame=none,framerule=0.1pt,framesep=6pt,numbers=left,xleftmargin=25pt,fontfamily=tt,fontsize=\small}
\newenvironment{hanging}
    {\begin{list}{}{\setlength\itemsep{0pt}%
 \setlength\topsep{0pt}%
 \setlength\leftmargin{25pt}%
 \setlength\itemindent{0pt}%
 \setlength\listparindent{\itemindent}}%
     \item[]}
    {\end{list}}

\newcommand{\toprule}{\par\vspace{1ex}\noindent\hspace{25pt}\rule{435pt}{.1pt}}
\newcommand{\botrule}{\noindent\hspace{25pt}\rule{435pt}{.1pt}\\[2ex]}

\newenvironment{cdbin}{\fvset{firstnumber=1}\color[named]{Blue}\Verbatim}{\endVerbatim}
\newenvironment{cdbcont}{\fvset{firstnumber=last}\color[named]{Blue}\Verbatim}{\endVerbatim}
\newenvironment{cdbout}{\vspace{-1.4ex}\begin{equation}}{\end{equation}\vspace{-1.4ex}}
\newenvironment{cdbcom}{\begin{hanging}}{\end{hanging}}
\newcommand{\Cdb}{{Cadabra}\xspace}

\begin{document}
\pagestyle{empty}
\begin{flushright}
hep-th/0701238\\
25 January 2007\\
3 April 2018 (update for v2 syntax)
\end{flushright}
\vskip 7ex

\begin{center}
\begin{minipage}{.95\textwidth}
{\huge\bf Introducing Cadabra: a symbolic computer\\[1ex]  
algebra system for field theory problems}\\[7ex]
{\large\bf Kasper Peeters}\\[5ex]
Department of Mathematical Sciences\\
Durham University\\
South Road\\
Durham DH1 3LE\\
United Kingdom\\[3ex]
{\tt kasper.peeters@durham.ac.uk}
\vskip 9ex

{\bf Abstract:}\\[1ex] Cadabra is a new computer algebra system
designed specifically for the solution of problems encountered in
field theory. It has extensive functionality for tensor polynomial
simplification taking care of Bianchi and Schouten identities, for
fermions and anti-commuting variables, Clifford algebras and Fierz
transformations, implicit coordinate dependence, multiple index types
and many other field theory related concepts. The input format is a
subset of \TeX{} and thus easy to learn. Both a command-line and a
graphical interface are available. The present paper is an
introduction to the program using several concrete problems from
gravity, supergravity and quantum field theory.
\end{minipage}
\vfill\vfill

\begin{minipage}{0.95\textwidth}
\noindent{\smaller\smaller This paper originally appeared in 2007. It
  has been updated to reflect the syntax changes introduced with
  Cadabra~2.x (first released in 2016), but is otherwise essentially
  unchanged. A more in-depth discussion of the changes of the 2.x
  series will appear in an upcoming paper. The software itself,
  including source code, can be obtained from the web site
  \url{http://cadabra.science/} where further documentation and
  tutorial notebooks can also be found.}
\end{minipage}
\end{center}
\eject
\pagestyle{plain}
\hrule
\tableofcontents
\vspace{4ex}
\hrule
\bigskip

\section{Introduction}

The currently available spectrum of symbolic computer algebra software
exhibits, from the perspective of a high-energy physicist, an
unfortunate deficiency. On the one hand, problems dealing with
differential equations, symbolic matrix algebra, special functions,
series expansions, polynomial algebra and so on are adequately
covered, often by several different systems. It is not too hard to
translate a give physics problem in these classes from paper to
computer and back. Other common problems, on the other hand, such as
applying symmetry transformations to Lagrangians, computing Poisson
brackets, deriving field equations, canonicalising tensor expressions
or performing Fierz transformations, are much harder to handle using
existing symbolic computer algebra tools. For these classes of problems,
which one could label as ``field theory problems'', it typically takes
much more effort to convert them from paper and solve them on the computer.

Granted, there are many smaller systems, often built on top of general
purpose systems such as Mathematica or Maple, which solve one or more
of these problems.\footnote{Among the more often-used recently written
packages there is {\tt GAMMA} for gamma matrix
algebra~\cite{Gran:2001yh}, {\tt grassmann.m} for handling of
anti-commuting variables~\cite{e_grassmann} and the extensive {\tt
xAct} for tensor manipulation~\cite{e_xact}. This list is far from
complete and only meant to illustrate how much work one has to do in
order to teach a general-purpose symbolic computer algebra system
about even the simplest concepts which occur in field theory.}
However, writing such packages, or even adapting them for slightly
more complicated situations than for which they were intended, is
often far from easy, and requires a substantial knowledge of the
underlying computer algebra system. Moreover, combining them, such
that e.g.~a tensor manipulating package can deal with fermions too, is
typically a tedious exercise, if possible at all. There are clear
technical reasons for this, rooted in the design of general purpose
computer algebra systems. However, rather than dwelling on these
deficiencies, it is more productive to actually come up with a better
solution.

The present paper is an introduction to a new computer algebra system,
called ``Cadabra'', which is designed from the start with field theory
problems in mind. By discussing a number of explicit sample problems
representative of the ``field theory'' class vaguely defined above,
this paper is intended as a guide for users rather than computer
scientists (for those interested, a description of the technical
implementation and code design goals has appeared
in~\cite{Peeters:2006kp}).

\medskip

In which sense does \Cdb differ from other computer algebra systems?
The first and most easily visible feature is that all expression input is in
the form of a subset of \TeX. Tensor indices, Dirac conjugation,
derivative operators, commutators, fermion products and so on are all
written just as in \TeX.  With a little bit of discipline, one can
cut-and-paste expressions straight from a paper into a \Cdb
notebook. The output, similarly, is typeset as \TeX{} would do it, and
\Cdb notebook files are in fact at the same time also valid \TeX{}
files (the program comes with a graphical notebook interface, but can
also be used from the command line).

Secondly, \Cdb contains some of the most powerful tensor expression
simplification routines available at present. Not only does it use
simple symmetry or anti-symmetry of tensors or tensor indices in order
to simplify expressions\footnote{I am grateful to Jos\'e
Mart\'\i{}n-Garc\'\i{}a for allowing me to use his excellent {\tt
xPerm} code~\cite{e_xact} for this purpose.}, it also takes into
account multi-term relations such as the Bianchi identity, as well as
dimension-dependent relations such as the Schouten identity
(section~\ref{s:bianchi}). The program handles commuting as well as
anti-commuting tensors, allows for multiple index sets
(section~\ref{s:example2}), and knows about the concept of a dummy
index so that no special wildcard notation is ever needed when
handling tensor indices.

Thirdly, the program knows about many concepts which are common in
field theory. It handles anti-commuting and non-commuting objects
without special notations for their products (section~\ref{s:other}),
it knows about gamma matrix algebra (section~\ref{s:gamma}), Fierz
identities (section~\ref{s:example3}), Dirac conjugation, vielbeine,
flat and curved, covariant and contravariant indices
(section~\ref{s:other}), implicit dependence of tensors on
coordinates, partial and covariant derivatives
(section~\ref{s:derivatives}). It has extensive facilities for handling
of field theory expressions, e.g.~dealing with variational
derivatives. It features a substitution command which correctly
handles anti-commuting objects and dummy indices and offers a wide
variety of pattern matching situations which occur in field theory.

Fourthly, the program has special support that allows the user to
write (and keep) expressions in any desired form, e.g.~to specify the
preferred order of objects. The program does not try to do anything
smart unless it is explicitly told to do so. However, when desired, it
is always possible to add any arbitrary simplification step to a list
of commands which is executed at every step of the calculation.  In
this way one has precise control over the level of verbosity of
the intermediate results of a computation.

Finally, the source code of the program is freely available, with
extensive documentation on how to extend it with new algorithms and
new data types (e.g.~twistor variables are planned for an upcoming
release). The program is completely independent of commercial software
and relies only on a few other open source libraries and programs. It
runs on Linux, Mac~OS~X, Windows and various BSD flavours and the
source code as well as binary packages can be downloaded from the web
site.

\medskip

The following sections discuss these characteristic features of \Cdb,
illustrated with many explicit calculations and several longer
examples (see in particular the Riemann tensor polynomial problem in
section~\ref{s:example1}, the Kaluza-Klein gravity problem in
section~\ref{s:example2} and the fermion Fierz problem in
section~\ref{s:example3}). These examples were chosen to illustrate
the typical use of \Cdb, in particular the way in which it fills a gap
in the existing spectrum of computer algebra software. The web site
offers a growing collection of longer sample calculations for those
readers who are interested in applications to more advanced problems.
The program is under active development and information about updates
and new features can also be obtained from there.

\section{Bosonic basics}
\subsection{Tensors, indices and symmetries}

Before discussing actual calculations, let us start with a few words
concerning notation. This discussion can be short because, as
mentioned in the introduction, mathematical expressions are entered
essentially as one would enter them in \TeX{} (with a few restrictions
to avoid ambiguities, which we will discuss as we go along). In order
to manipulate expressions, \Cdb often needs to know a bit more about
properties of tensors or other symbols. Such properties are entered
using the `\verb|::|' symbol. A simple example is the declaration of
index sets, useful for automatic dummy index relabelling. An example
will clarify this,\footnote{The input and output shown in this paper
are essentially identical to those in the graphical interface which
comes with \Cdb, the only difference being the added line numbers and
some minor typographical modifications.}  \toprule
\begin{cdbin}
{ a, b, c, d }::Indices.
ex:= A_{a b} B_{b c};
\end{cdbin}
\begin{cdbout}
A_{a b} B_{b c};
\end{cdbout}
\begin{cdbcont}
substitute(_, $B_{a b} ->  C_{a b c} D_{c}$ );
\end{cdbcont}
\begin{cdbout}
A_{a b} \, C_{b c d} \, D_{d};
\end{cdbout}
\botrule
The automatic index relabelling which has taken place in this
substitute command was clearly only possible because of the property
declaration in the first line. Note how the substitute command has
also figured out that \verb|B_{a b}| on the left-hand side is
equivalent to \verb|B_{b c}|, without any explicit wildcards or
patterns. We will see more of this type of field-theory motivated
logic throughout the paper.

Indices can be simple letters, as in the example above, but it is also
perfectly possible to put accents on them. This can be useful for
e.g.~notations involving chiral spinors. The following example
illustrates the use of accents on indices.
\toprule
\begin{cdbin}
A_{\dot{a} \dot{b}}::AntiSymmetric.
ex:= A_{\dot{b} \dot{a}};
\end{cdbin}
\begin{cdbout}
A_{\dot{b} \dot{a}};
\end{cdbout}
\begin{cdbcont}
canonicalise(_);
\end{cdbcont}
\begin{cdbout}
(-1)\,A_{\dot{a} \dot{b}};
\end{cdbout}
\botrule Here we also see a second usage of property declarations: the
construction in the first line declares that the $A_{\dot{a} \dot{b}}$
tensor is antisymmetric in its indices. The canonicalise command
subsequently writes the input in a canonical form, which in this
trivial example simply means that the indices gets sorted in
alphabetical order.  Continuing the example above, one can also use
subscripts or superscripts on indices, as in the example below.
\vfill\eject
\toprule
\begin{cdbin}
{ a_{1}, a_{2}, a_{3}, a_{4} }::Indices(vector).
ex:= V_{a_{1}} W_{a_{1}}:
substitute(_, $V_{a_{2}} -> M_{a_{2} a_{1}} N_{a_{1}}$ );
\end{cdbin}
\begin{cdbout}
M_{a_{1} a_{2}} \, N_{a_{2}} \, W_{a_{1}};
\end{cdbout}
\botrule As this example shows in some more detail, the input format
is a mixture of Cadabra's own LaTeX-like language for the description
of mathematical expressions, and Python. The underscore symbol
`\verb|_|' always refers to the last-used expression.

A guiding principle in \Cdb is that nothing ever has to be declared
unless this is absolutely needed. This is in contrast to many other
systems, where for instance one has to declare manifolds and index
sets and so on before one can even enter a tensor expression. This
makes code hard to read, but more importantly, such additional
declarations are hard to remember. As an example of how \Cdb works,
one can e.g.~input tensor expressions and perform substitution on them,
without ever declaring the symbols used for indices. Only when the
program needs to generate new dummy indices does one need to declare
index sets, so that dummy indices can be taken out of the right
set. The general guideline is that ``one only
needs to tell the program about the meaning of symbols when this is
actually required to do the manipulation correctly''.

In order to declare symmetries of tensors, it is possible to use
simple shorthands like the \verb|AntiSymmetric| in the example above. 
More generally, symmetries can be declared using  a generic Young
tableau notation. An object with the symmetries of a Riemann tensor,
for example, can be declared as in the following example.
\toprule
\begin{cdbin}
R_{a b c d}::TableauSymmetry(shape={2,2}, indices={0,2,1,3}).
ex:=R_{a b c d} R_{d c a b}:
canonicalise(_);
\end{cdbin}
\begin{cdbout}
(-1)\, R_{a b c d}\, R_{a b c d};
\end{cdbout}
\botrule
The first line indicates that the tensor has the symmetries of the
\raisebox{2ex}{$\ftableau{ac,bd}$} tableau (the numbers in the \verb|indices| argument
refer to the index positions). The \verb|canonicalise| algorithm writes
the input in canonical form with respect to mono-term symmetries
(anti-symmetry in the two index pairs and symmetry under pair
exchange). The Ricci symmetry is also encoded in the Young tableau and
will be discussed later. Many tensor symmetries have shorthand
notations, so one often does not have spell out the tableau (e.g.~\verb|RiemannTensor| or \verb|SatisfiesBianchi|).

\subsection{Derivatives and dependencies}
\label{s:derivatives}

There are relatively few surprises when it comes to derivatives.
It is possible to write derivatives with respect to coordinates,
i.e.~write~$\partial_x$ where~$x$ is a coordinate, but
it is also possible to use indices, as in~$\partial_i$ with $i$ being
a vector index. It is also possible to make objects implicitly
dependent on a derivative operator, so that one can write~$\partial
A$ without an explicit specification of the coordinate which is
involved here. 

In order to make this possible, however, derivative objects have to be
declared just like indices, otherwise the system does not know which
symbol ($\partial$, $D$, $d$, $\nabla$, \ldots) one wants to use for
them. Implicit dependencies of objects on coordinates associated to
derivatives is indicated through a \verb|Depends| property. Here is an
example to illustrate all this: \toprule
\begin{cdbin}
\nabla{#}::Derivative.
\partial{#}::PartialDerivative.
A_{m n}::AntiSymmetric.
V_{m}::Depends(\nabla{#}).

ex:= \partial_{m p}( A_{q r} V_{n} ) A^{p m};
\end{cdbin}
\begin{cdbout}
\partial_{m p}( A_{q r} V_{n} ) A^{p m};
\end{cdbout}
\begin{cdbcont}
canonicalise(_);
\end{cdbcont}
\begin{cdbout}
0;
\end{cdbout}
\begin{cdbcont}
ex:=\nabla_{m p}( A_{q r} V_{n} ) A^{p m};
canonicalise(_);
\end{cdbcont}
\begin{cdbout}
(-1) \nabla_{m p}( A_{q r} V_{n} ) A^{m p};
\end{cdbout}
\begin{cdbcont}
unwrap(_);
\end{cdbcont}
\begin{cdbout}
(-1) A_{q r} \nabla_{m p}{ V_{n} } A^{m p};
\end{cdbout}
\botrule
Note how the symmetry of a double partial derivative has automatically
been taken into account (it
is part of the \verb|PartialDerivative| property). This is called
``property inheritance''.

\subsection{Bianchi, Ricci and Schouten identities}
\label{s:bianchi}

So far we have seen several examples of so-called ``mono-term''
canonicalisation, in which simple symmetries of tensors are used,
which relate one particular term to one particular other term. More
complicated symmetries are Ricci or Bianchi identities, which relate
more than two terms (``multi-term'' symmetries). Such identities can
be taken into account using Young tableau projectors. Here is an
example to show how this works.  
\toprule
\begin{cdbin}
{m,n,p,q,r,s,t#}::Indices(vector).
\nabla{#}::Derivative.
R_{m n p q}::RiemannTensor.
\nabla_{m}{R_{p q r s}}::SatisfiesBianchi.
\end{cdbin}
\begin{cdbcom}
The last line is a shorthand, but internally does nothing more than to
associate a particular Young tableau symmetry to the given
tensor.\footnote{The alternative is
{\tt $\backslash$nabla\_\{m\}\{R\_\{p q r s\}\}::TableauSymmetry( shape=\{3,2\}, indices=\{1,3,0,2,4\} ).} 
which makes use of the more general tableau symmetry property which we
have seen earlier.} Here it effectively associates the $\nabla$
operator to the Riemann tensor. We can see this in action, for
instance, by verifying that the Bianchi identity indeed holds,
\footnote{The {\tt depth=1} parameter of the
  {\tt young\_project\_tensor} algorithm indicates that this command
  should only be applied at ``level 1'' of the expression, i.e.~at the
  level of the terms of the sum.}
\end{cdbcom}
\begin{cdbcont}
ex:= \nabla_{m}{R_{p q r s}} + \nabla_{p}{R_{q m r s}} + \nabla_{q}{R_{m p r s}}:
young_project_tensor(_, depth=1, modulo_monoterm=True);
\end{cdbcont}
\begin{cdbout}
0
\end{cdbout}
\botrule
As expected, by Young-projecting the expression, the Bianchi identity
becomes manifest~\cite{Green:2005qr}.

A similar logic can also be used~\cite{kas_dimdep} to take into
account dimension-dependent identities, more generally known as
Schouten identities or Lovelock identities. Take as an example two
anti-symmetric tensors~$A_{m n p}$ and~$B_{m n p}$. Then we have
\begin{equation}
A_{m n p} B_{m n q} - A_{m n q} B_{m n p} = 0 \quad\text{in $d=3$,}
\end{equation}
which is conventionally proved by using the fact that
anti-symmetrising in four indices yields zero in three dimensions.
\Cdb instead uses Young tableau product rules to accomplish the same
result~\cite{kas_dimdep}. 
\toprule
\begin{cdbin}
{ m, n, p, q }::Indices(vector).
{ A_{m n p}, B_{m n p} }::AntiSymmetric.
A_{m n p} B_{m n q} - A_{m n q} B_{m n p};
\end{cdbin}
\begin{cdbout}
ex:= A_{m n p} B_{m n q} - A_{m n q} B_{m n p};
\end{cdbout}
\begin{cdbcom}
In four dimensions or higher, this product cannot be simplified any
further. Decomposing the product using Young projectors and
canonicalising with respect to monoterm symmetries gives back the input,
\end{cdbcom}
\begin{cdbcont}
{ m, n, p, q }::Integer(1..4).
decompose_product(_)
canonicalise(_);
\end{cdbcont}
\begin{cdbout}
A_{p m n} B_{q m n} - A_{q m n} B_{p m n};
\end{cdbout}
\begin{cdbcom}
In three dimensions, however, the expression vanishes by virtue of the
Schouten identity. The same procedure as above now indeed produces a zero,
\end{cdbcom}
\begin{cdbcont}
{ m, n, p, q }::Integer(1..3).
decompose_product(_)
canonicalise(_);
\end{cdbcont}
\begin{cdbout}
0; 
\end{cdbout}
\botrule
Some more examples can be found in~\cite{Peeters:2006kp} and in the reference manual.

As an added bonus, the Young tableau routines which are used
internally can also be used directly. There are properties both for
filled and unfilled tableaux. The following example shows an example
of the latter, together with the Littlewood-Richardson algorithm and
the use of the hook formula to compute dimensions of representations,
\eject

\toprule
\begin{cdbin}
\ftableau{#}::FilledTableau(dimension=10).
ex:=\ftableau{0,0}{1,1} \ftableau{a,a}{b,b}:
lr_tensor(_);
\end{cdbin}
\begin{cdbout}
\ftableau{{0}{0}{a}{a},{1}{1}{b}{b}} \oplus \ftableau{{0}{0}{a}{a},{1}{1}{b},{b}} \oplus \ftableau{{0}{0}{a}{a},{1}{1},{b}{b}} \oplus \ftableau{{0}{0}{a},{1}{1}{b},{a},{b}} \oplus \ftableau{{0}{0}{a},{1}{1},{a}{b},{b}} \oplus \ftableau{{0}{0},{1}{1},{a}{a},{b}{b}};
\end{cdbout}
\begin{cdbcont}
tab_dimension(_);
\end{cdbcont}
\begin{cdbout}
70785 \oplus 235950 \oplus 154440 \oplus 101640 \oplus 103950 \oplus 13860;
\end{cdbout}
\botrule

\subsection{A Riemann tensor polynomial example}
\label{s:example1}

Let us now discuss a somewhat more complicated example of bosonic
tensor algebra, which shows not only how symmetries of tensors are
used, but also how repeated derivative operators work and how
more complicated canonicalisation problems are solved.

To this extent we will discuss a certain identity, a
proof of which can be found in appendix~A of~\cite{Frolov:2001jh}.
Concretely, the problem concerns two particular third-order
polynomials in the Weyl tensor~$C_{m n p q}$, given by
\begin{equation}
\begin{aligned}
E_{ij} &= - C_{i}{}^{mkl} C_{jpkq} C_{l}{}^{pmq}
+ \frac{1}{4} C_{i}{}^{mkl} C_{jmpq} C_{kl}{}^{pq}
- \frac{1}{2} C_{ikjl} C^{kmpq} C^{l}{}_{mpq}\,,\\[1ex]
E &= C_{jmnk} C^{mpqn} C_{p}{}^{jk}{}_q + \frac{1}{2} C_{jkmn} C^{pqmn} C^{jk}{}_{pq}\,.
\end{aligned}
\end{equation}
When evaluated on an Einstein space, these polynomials are supposed to
satisfy the identity
\begin{equation}
\label{e:Frolovid}
\nabla_i \nabla_j E_{ij} - \frac{1}{6} \nabla_i \nabla_i E = 0\,.
\end{equation}
Proving this is a tedious exercise with Bianchi identities when done by
hand. With \Cdb the proof is straightforward; the notebook is depicted
below.

The key elements in this example are the declaration of the Weyl
tensor as well as the covariant derivative, together with the 4th line
which links the two, by stating that the covariant derivative of the
Weyl tensor satisfies the Bianchi identity. 
\vfill\eject

\toprule
\begin{cdbin}
{i,j,m,n,k,p,q,l,r,r#}::Indices(vector).
C_{m n p q}::WeylTensor.
\nabla{#}::Derivative.
\nabla_{r}{ C_{m n p q} }::SatisfiesBianchi.

Eij:=- C_{i m k l} C_{j p k q} C_{l p m q} + 1/4 C_{i m k l} C_{j m p q} C_{k l p q}
     - 1/2 C_{i k j l} C_{k m p q} C_{l m p q}:

E:=  C_{j m n k} C_{m p q n} C_{p j k q} + 1/2 C_{j k m n} C_{p q m n} C_{j k p q}:

exp:= \nabla_{i}{\nabla_{j}{ @(Eij) }} - 1/6 \nabla_{i}{\nabla_{i}{ @(E) }}:
\end{cdbin}
\begin{cdbcom}
We now need to apply (twice) the Leibniz rule to expand the
derivatives, and then sort the tensors and write the result in
canonical form with respect to mono-term symmetries,
\end{cdbcom}
\begin{cdbcont}
distribute(_); product_rule(_);
distribute(_); product_rule(_);  

sort_product(_); canonicalise(_);
rename_dummies(_);
\end{cdbcont}
\begin{cdbcom}
Because the identity which we intend to prove is only supposed to hold
on Einstein spaces, we set the divergence of the Weyl tensor to zero,
\end{cdbcom}
\begin{cdbcont}
substitute(_, $\nabla_{i}{C_{k i l m}} -> 0 , \nabla_{i}{C_{k m l i}} -> 0$ );
\end{cdbcont}
\begin{cdbout}
\begin{aligned}
& ~~~~ C_{i j m n} C_{i k m p} \nabla_{q}{\nabla_{j}{C_{n k p q}}} 
- C_{i j m n} \nabla_{k}{C_{i p m q}} \nabla_{p}{C_{j q n k}} 
- 2 C_{i j m n} \nabla_{i}{C_{m k p q}} \nabla_{p}{C_{j k n q}} \\
&- C_{i j m n} C_{i k p q} \nabla_{m}{\nabla_{p}{C_{j q n k}}} 
- \frac{1}{4} C_{i j m n} C_{i j k p} \nabla_{q}{\nabla_{m}{C_{n q k p}}} 
+ \frac{1}{4} C_{i j m n} \nabla_{k}{C_{i j p q}} \nabla_{p}{C_{m n k q}} \\
&- \frac{1}{2} C_{i j m n} \nabla_{i}{C_{j k p q}} \nabla_{k}{C_{m n p q}} 
+ \frac{1}{4} C_{i j m n} C_{i k p q} \nabla_{j}{\nabla_{k}{C_{m n p q}}} 
+ \frac{1}{2} C_{i j m n} C_{i k p q} \nabla_{m}{\nabla_{j}{C_{n k p q}}} \\
&+ \frac{1}{2} C_{i j m n} \nabla_{i}{C_{m k p q}} \nabla_{n}{C_{j k p q}} 
- \frac{1}{2} C_{i j m n} \nabla_{i}{C_{j k p q}} \nabla_{m}{C_{n k p q}} 
+ \frac{1}{2} C_{i j m n} C_{i k p q} \nabla_{j}{\nabla_{m}{C_{n k p q}}}\\ 
&+ \frac{1}{2} C_{i j m n} C_{i k m p} \nabla_{q}{\nabla_{q}{C_{j k n p}}} 
+ C_{i j m n} \nabla_{k}{C_{i p m q}} \nabla_{k}{C_{j p n q}} 
- \frac{1}{4} C_{i j m n} C_{i j k p} \nabla_{q}{\nabla_{q}{C_{m n k p}}} \\
&- \frac{1}{2} C_{i j m n} \nabla_{k}{C_{i j p q}} \nabla_{k}{C_{m n p q}};
\end{aligned}
\end{cdbout}
\begin{cdbcom}
This expression should vanish upon use of the Bianchi identity. By
expanding all tensors using their Young projectors, this becomes manifest,
\end{cdbcom}
\begin{cdbcont}
young_project_product(_);
\end{cdbcont}
\begin{cdbout}
0;
\end{cdbout}
\begin{cdbcom}
This proves the identity~\eqref{e:Frolovid}.
\end{cdbcom}
\botrule
The use of Young projector methods also allows for other calculations
which are computationally intensive when done by hand. An example is
the problem of generating a complete basis of monomials of Riemann
tensors, discussed in~\cite{Peeters:2006kp}. 


\subsection{Index ranges and subspaces, Kaluza-Klein gravity}
\label{s:example2}

We have so far seen rather simple uses of indices, in which only a
single vector space was used. \Cdb contains functionality to deal with
more complicated situations though. In section~\ref{s:fermions} we
will discuss the use of multiple index types as well as implicit
indices. Here we will discuss another index feature, namely the
possibility to ``split'' indices into two or more subspaces. Let us
first illustrate this with a simple example.
\toprule
\begin{cdbin}
{M, N, P}::Indices(space).
{m, n, p}::Indices(subspace1).
{a, b, c}::Indices(subspace2).

ex:= A_{M N} B_{N P};
split_index(_, $M, m, a$);
\end{cdbin}
\begin{cdbout}
A_{M m} \, B_{m P} + A_{M a}\, B_{a P}\,;
\end{cdbout}
\botrule 
The first three lines declare three types of indices, labelled as
``space'', ``subspace1'' and ``subspace2'' respectively. The last line
of the input performs the split of the dummy index into the two
subspaces. Instead of using two subspaces labelled by indices, it is
also possible to use a one-dimensional subspace. An example of how
this works can be found in the Kaluza-Klein problem discussed below.

The index split functionality is particularly useful for Kaluza-Klein
type problems, in which the metric is decomposed according to
\begin{equation}
\label{e:KKansatz}
g_{\mu\nu} = \begin{pmatrix}
 \phi^{-1}\, h_{m n} + \phi\, A_{m} A_{n}  & \phi\, A_{m} \\
\phi\, A_{n} & \phi 
\end{pmatrix}\,,
\end{equation}
(where the usual conventions in four space-time dimensions were used).
It is a somewhat tedious exercise to compute the Riemann tensor
components for this particular metric ansatz. One finds for instance
that 
\begin{equation}
\begin{aligned}
\label{e:Rm4n4}
R_{m 4 n 4} &=  - \frac{1}{2}  \nabla_{m}\partial_{n}{\phi} 
 - \frac{1}{4} \partial_{m}{\phi} \partial_{n}{\phi}\, \phi^{-1} 
 + \frac{1}{4} \partial_{p}{\phi} \partial_{q}{\phi} \phi^{-1} h_{m n} h^{p q} 
 + \frac{1}{4} F_{m p} F_{n q} \phi^{3} h^{p q} \,,
\end{aligned}
\end{equation}
where $F_{mn}$ is the field strength of~$A_{m}$.
The following notebook computes this expression using \Cdb and the
index splitting functionality. All intermediate output is suppressed
as it tends to get rather lengthy.
\toprule
\begin{cdbin}
{\mu,\nu,\rho,\sigma,\kappa,\lambda,\eta,\chi#}::Indices(full, position=fixed).
{m,n,p,q,r,s,t,u,v,m#}::Indices(subspace, position=fixed, parent=full).
\end{cdbin}
\begin{cdbcom}
Note the appearance of \verb|parent=full|. This indicates that the
indices in the second set span a subspace of the indices in the first
set. The remaining declarations are standard,
\end{cdbcom}
\begin{cdbcont}
\partial{#}::PartialDerivative.
g_{\mu\nu}::Metric.
g^{\mu\nu}::InverseMetric.
g_{\mu? \nu?}::Symmetric.
g^{\mu? \nu?}::Symmetric.
h_{m n}::Metric.
h^{m n}::InverseMetric.
\delta^{\mu?}_{\nu?}::KroneckerDelta.
\delta_{\mu?}^{\nu?}::KroneckerDelta.
F_{m n}::AntiSymmetric.
\end{cdbcont}
\begin{cdbcom}
We will want to expand the Riemann tensor in terms of the metric. The
following two substitution rules do the conversion from Riemann tensor
to Christoffel symbol and from Christoffel symbol to
metric.\footnote{Cadabra 2.x contains a growing library of packages
  with expressions of this type, but we will here not rely on that
  library facility.} Index patterns like \verb|\lambda?| match both four- and
three-dimensional indices.
\end{cdbcom}
\begin{cdbcont}
RtoG:= R^{\lambda?}_{\mu?\nu?\kappa?} -> 
 - \partial_{\kappa?}{ \Gamma^{\lambda?}_{\mu?\nu?} }
 + \partial_{\nu?}{ \Gamma^{\lambda?}_{\mu?\kappa?} }
 - \Gamma^{\eta}_{\mu?\nu?} \Gamma^{\lambda?}_{\kappa?\eta}
 + \Gamma^{\eta}_{\mu?\kappa?} \Gamma^{\lambda?}_{\nu?\eta}:

Gtog:= \Gamma^{\lambda?}_{\mu?\nu?} ->
  (1/2) * g^{\lambda?\kappa} ( 
        \partial_{\nu?}{ g_{\kappa\mu?} } + \partial_{\mu?}{g_{\kappa\nu?} } 
                                       - \partial_{\kappa}{ g_{\mu?\nu?} } ):
\end{cdbcont}
\begin{cdbcom}
Now input the $R_{m 4 n 4}$ component and do the substitution. After
each substitution, we distribute products over sums. We also apply the
product rule to distribute derivatives over factors in a product.
\end{cdbcom}
\begin{cdbcont}
todo:= g_{m1 m} R^{m1}_{4 n 4} + g_{4 m} R^{4}_{4 n 4};
substitute(_, RtoG)
substitute(_, Gtog)
distribute(_)
product_rule(_)
distribute(_)
sort_product(_)
\end{cdbcont}
\begin{cdbcom}
We now split the $\mu$ index into a $m$ part and the remaining $4$
direction (the \verb|!!| version of the command makes it apply until
the result no longer changes).  After that, we remove $x^4$ derivatives of the gauge
field and write the expression in canonical form,
\end{cdbcom}
\begin{cdbcont}
split_index(_, $\mu, m1, 4$, repeat=True)
substitute(_, $\partial_{4}{A??} -> 0$, repeat=True)
substitute(_, $\partial_{4 m?}{A??} -> 0$, repeat=True)
substitute(_, $\partial_{m? 4}{A??} -> 0$, repeat=True)
canonicalise(_);
\end{cdbcont}
\begin{cdbcom}
In the next step, we insert the metric ansatz~\eqref{e:KKansatz} and
simplify the result as much as possible.
\end{cdbcom}
\begin{cdbcont}
substitute(_, $g_{4 4} -> \phi$ )
substitute(_, $g_{m 4} -> \phi A_{m}$ )
substitute(_, $g_{4 m} -> \phi A_{m}$ )
substitute(_, $g_{m n} -> \phi**{-1} h_{m n} + \phi A_{m} A_{n}$ )
substitute(_, $g^{4 4} -> \phi**{-1} + \phi A_{m} h^{m n} A_{n}$ )
substitute(_, $g^{m 4} -> - \phi h^{m n} A_{n}$ )
substitute(_, $g^{4 m} -> - \phi h^{m n} A_{n}$ )
substitute(_, $g^{m n} -> \phi h^{m n}$ );
\end{cdbcont}
\begin{cdbcom}
Some derivatives have to be rewritten to a canonical form,
\end{cdbcom}
\begin{cdbcont}
converge(todo):
   distribute(_)
   product_rule(_)
   canonicalise(_)
\end{cdbcont}
\begin{cdbcom}
The above shows the use of a Cadabra extension \verb|converge|, which
applies a list of algorithms to an expression until it no longer
changes. We now rewrite derivatives of inverse metrics,
\end{cdbcom}
\begin{cdbcont}
substitute(_, $\partial_{p}{h^{n m}} h_{q m} -> - \partial_{p}{h_{q m}} h^{n m}$ )
collect_factors(_)
sort_product(_)
converge(todo):
   substitute(_, $h_{m1 m2} h^{m3 m2} -> \delta_{m1}^{m3}$, repeat=True )
   eliminate_kronecker(_)
   canonicalise(_)
\end{cdbcont}
\begin{cdbcom}
Finally, we replace the derivative of the gauge field with the field
strength,
\end{cdbcom}
\begin{cdbcont}
substitute(_, $\partial_{n}{A_{m}} -> 1/2*\partial_{n}{A_{m}}
               + 1/2*F_{n m} + 1/2*\partial_{m}{A_{n}}$ )
distribute(_)
sort_product(_)
canonicalise(_)
rename_dummies(_);
\end{cdbcont}
\begin{cdbout}
\begin{aligned}
{}& - \frac{1}{4} \partial_{m}{\phi} \partial_{n}{\phi} \phi^{-1} 
 + \frac{1}{4} \partial_{p}{\phi} \partial_{n}{h_{m q}} h^{p q} 
 - \frac{1}{2} \partial_{m n}{\phi} 
 + \frac{1}{4} F_{m p} F_{n q} \phi^{3} h^{p q} \\
{}& + \frac{1}{4} \partial_{p}{\phi} \partial_{q}{\phi} \phi^{-1} h_{m n} h^{p q} 
 - \frac{1}{4} \partial_{p}{\phi} \partial_{q}{h_{m n}} h^{p q} 
 + \frac{1}{4} \partial_{p}{\phi} \partial_{m}{h_{n q}} h^{p q};
\end{aligned}
\end{cdbout}
\begin{cdbcom}
This is indeed equivalent to~\eqref{e:Rm4n4} upon writing out the
covariant derivative in the first term of that equation.
\end{cdbcom}
\botrule
If required, some of these calculations can be done with fewer lines
of input by adding a number of default simplification rules; an
example of such default rules will be discussed in the next section. 
We will end here the discussion of purely bosonic problems and move on
to fermions and anti-commuting tensors.

\vfill\eject
\section{Fermions, Dirac algebra and Fierz transformations}
\label{s:fermions}
\subsection{Simple gamma matrix algebra}
\label{s:gamma}

\Cdb has built-in algorithms for the manipulation of anti-commuting
objects, spinors and gamma matrices in any dimension. Combined with
the option of ``suppressing'' indices (in our examples below we will
suppress spinor indices), it becomes possible to write calculations in
a natural and compact way. All anti-commuting and fermionic objects as
usual need to be declared by attaching the appropriate properties to
them; we will see many examples of this.

Let us start, however, with some simple gamma matrix algebra. As an
example, we will expand the product $\Gamma_{s r} \Gamma_{r l}
\Gamma_{k m} \Gamma_{m s}$ in arbitrary dimensions in terms of the
irreducible~$\Gamma_{kl}$ and $\delta_{kl}$ components.  
\toprule
\begin{cdbcom}
We first declare the vector indices, their range, and the symbols used for
gamma matrices and Kronecker deltas.
\end{cdbcom}
\begin{cdbcont}
{s,r,l,k,m,n}::Indices(vector).
{s,r,l,k,m,n}::Integer(0..d-1).
\Gamma_{#}::GammaMatrix(metric=\delta).
\delta_{m n}::KroneckerDelta.
\end{cdbcont}
\begin{cdbcom}
The declaration for the gamma matrix shows that we are defining an
object with implicit indices: the spinor indices will be
suppressed. The notation \verb|_{#}| denotes the presence of an
arbitrary number of indices.

It is useful to let Cadabra do a bit more simplification at every step
(more than the default of simply collecting equal terms). This can be
achieved by re-defining the \verb|post_process| function, which gets
called after every step of the computation.
\end{cdbcom}
\begin{cdbin}
def post_process(ex):
   sort_product(ex)
   eliminate_kronecker(ex)
   canonicalise(ex)
   collect_terms(ex)
\end{cdbin}
\begin{cdbcom}
This sorts the product, eliminate Kronecker delta's, writes indices in
canonical order and then finally collects equal terms.
  
Next follows the actual computation. We write down the gamma matrix
product and join gamma matrices three times,
\end{cdbcom}
\begin{cdbcont}
ex:= \Gamma_{s r} \Gamma_{r l} \Gamma_{k m} \Gamma_{m s};
for i in range(3):
   join_gamma(_)
   distribute(_)
\end{cdbcont}
\begin{cdbcom}
  Note once more how we are mixing ordinary Python loop constructions
  with Cadabra code here. 
\end{cdbcom}
\begin{cdbout}
\label{e:gam1}
-18 \Gamma_{kl} d + 8 \Gamma_{kl} d d + 12 \Gamma_{kl} - 3\delta_{kl}
+ 6 \delta_{kl} d - 4 \delta_{kl} d d - \Gamma_{k l} d d d +
\delta_{kl} d d d
\end{cdbout}
\begin{cdbcont}
factor_in(_, $d$)
collect_factors(_)
\end{cdbcont}
\begin{cdbout}
\Gamma_{k l} (- 18 d + 8 d^2 + 12 - d^3) + \delta_{k l} ( - 3 + 6 d - 4 d^2 + d^3);
\end{cdbout}
\botrule 
The key ingredient here is the \verb|join_gamma| algorithm, which
takes two adjacent generalised gamma matrices and expands their product in terms of
a basis of fully antisymmetrised gamma matrices. In the step
before~\eqref{e:gam1} the Python loop performs such a join three times
and expands out the resulting product. 

In the example above, the spinor indices on the gamma matrices were
suppressed. The \verb|GammaMatrix| property has turned the gamma
symbols into non-commuting objects, which will not change order when
sorting symbols in a product. It is, however, possible to add the
spinor indices back in, and use explicit indices. For this purpose,
\Cdb knows the concept of an ``index bracket'', which associates
indices to matrix objects like the gamma matrices above, or to
e.g.~vectors or spinors. Here is a somewhat simpler example:
\toprule
\begin{cdbin}
{a,b,c,d#}::Indices(spinor).
\Gamma_{#}::GammaMatrix(metric=\delta).
(\Gamma_{m n})_{a b} (\Gamma_{n p})_{b c};
combine(_);
\end{cdbin}
\begin{cdbout}
(\Gamma_{m n}\, \Gamma_{n p})_{a c};
\end{cdbout}
\begin{cdbcont}
join_gamma(_)
canonicalise(_);
\end{cdbcont}
\begin{cdbout}
 (\Gamma_{m p} \delta_{n n} - \Gamma_{m n} \delta_{n p} + \Gamma_{p n} \delta_{m n} 
+ \delta_{m p} \delta_{n n} - \delta_{m n} \delta_{n p})_{a c};
\end{cdbout}
\botrule
The join algorithm has acted `inside' the index bracket, on the matrix
objects themselves. Conversely, index brackets can
be distributed over elements in the sum, and objects can be taken out
of the index bracket when they are known not to contain implicit
indices:
\toprule
\begin{cdbcont}
distribute(_)
expand(_);
\end{cdbcont}
\begin{cdbout}
 (\Gamma_{m p})_{a c} \delta_{n n} - (\Gamma_{m n})_{a c} \delta_{n p} 
+ (\Gamma_{p n})_{a c} \delta_{m n} + (\delta_{m p} \delta_{n n})_{a c} 
- (\delta_{m n} \delta_{n p})_{a c};
\end{cdbout}
\botrule
In this way, matrix operations can be switched between abstract and
index notation at will.

\subsection{Fierz transformations}
\label{s:example3}

\Cdb can apply Fierz transformations in any dimension to re-order
four-fermion terms. As an example, consider the following identity for
Majorana spinors in eleven dimensions~\cite{deWit:1998tk},
\begin{equation}
\begin{aligned}
\label{e:fierzthing}
-e_{[\nu}{}^s (\bar \theta \Gamma^{rs} \psi_\rho)(\bar\psi_{\mu]} \Gamma^r \epsilon) &=
\sum_{n} \frac{1}{2^5\; n!} (\bar\psi_\mu \Gamma^{i_1\ldots i_n} \psi_\rho)
(\bar\theta \Gamma^{rs} \Gamma^{i_n\ldots i_1} \Gamma^{r} \epsilon) \\
&= \frac{1}{2^5} \bar\psi_{[\mu} \Gamma_m \psi_\rho e_{\nu]}{}^s \left( 8 \bar\theta \Gamma^{sm}\epsilon
+ 10 \eta^{sm} \bar\theta\epsilon\right) \\
&\quad - \frac{1}{2^5\; 2!} \bar\psi_{[\mu}\Gamma_{mn} \psi_\rho e_{\nu]}{}^s\left(
-6 \bar\theta\Gamma^{smn} \epsilon + 16\bar\theta \Gamma^{[m} \epsilon \eta^{n]s}\right)\\
&\quad + \frac{1}{2^5\; 5!} \bar\psi_{[\mu}\Gamma_{mnopq} \psi_\rho e_{\nu]}{}^s \left(
10 \eta^{s[m} \bar\theta\Gamma^{nopq]} \epsilon\right) \, .
\end{aligned}
\end{equation}
With conventional tools, it would take some time to convert this
problem to the computer. Proving this with \Cdb, on the other hand,
requires little more than defining symbols properly and inputting the
expression on the left-hand side as one would type it in a
paper. Moreover, reading off the output is simple as well, because \Cdb produces
output which is virtually identical to the right-hand side of the
equation above. The notebook with comments is displayed below.
\toprule
\begin{cdbin}
{\mu,\nu,\rho}::Indices(curved, position=fixed).
{m,n,p,q,r,s,t,u,v}::Indices(flat, position=independent).
{m,n,p,q,r,s,t,u,v}::Integer(0..10).
T^{#{\mu}}::AntiSymmetric.
\psi_{\mu}::SelfAntiCommuting.
\psi_{\mu}::Spinor(dimension=11, type=Majorana).
\theta::Spinor(dimension=11, type=Majorana).
\epsilon::Spinor(dimension=11, type=Majorana).
{\theta,\epsilon,\psi_{\mu}}::AntiCommuting.
\bar{#}::DiracBar.
\delta^{m n}::KroneckerDelta.
\Gamma^{#{m}}::GammaMatrix(metric=\delta).\end{cdbin}
\begin{cdbcom}
These lines define the properties of all the symbols. We now
input (minus) the left-hand side of~\eqref{e:fierzthing}, and do a Fierz
transformation to bring the~$\theta$ and~$\epsilon$ spinors together,
\end{cdbcom}
\begin{cdbcont}
ex:= T^{\mu\nu\rho} e_{\nu}^{s}
     \bar{\theta} \Gamma^{r s} \psi_{\rho}
     \bar{\psi_{\mu}} \Gamma^{r} \epsilon;

fierz(ex, $\theta, \epsilon, \psi_{\mu}, \psi_{\rho}$ );
\end{cdbcont}
\begin{cdbout}
\begin{aligned}
&{}- \frac{1}{32} T^{\mu \nu \rho} e_{\nu}^{s} \bar{\theta} \Gamma^{r s}
\Gamma^{r} \epsilon \bar{\psi_{\mu}} \psi_{\rho} 
{}- \frac{1}{32} T^{\mu \nu \rho} e_{\nu}^{s} \bar{\theta} \Gamma^{r s}
\Gamma^{m} \Gamma^{r} \epsilon \bar{\psi_{\mu}} \Gamma_{m} \psi_{\rho} \\[1ex]
&{}- \frac{1}{64} T^{\mu \nu \rho} e_{\nu}^{s} \bar{\theta} \Gamma^{r s}
\Gamma^{m n} \Gamma^{r} \epsilon \bar{\psi_{\mu}} \Gamma_{n m}
\psi_{\rho} 
{}- \frac{1}{192} T^{\mu \nu \rho} e_{\nu}^{s} \bar{\theta} \Gamma^{r s}
\Gamma^{m n p} \Gamma^{r} \epsilon \bar{\psi_{\mu}} \Gamma_{p n m}
\psi_{\rho} \\[1ex]
&{}- \frac{1}{768} T^{\mu \nu \rho} e_{\nu}^{s} \bar{\theta} \Gamma^{r s}
\Gamma^{m n p q} \Gamma^{r} \epsilon \bar{\psi_{\mu}} \Gamma_{q p n m}
\psi_{\rho} 
{}- \frac{1}{3840} T^{\mu \nu \rho} e_{\nu}^{s} \bar{\theta} \Gamma^{r s}
\Gamma^{m n p q t1} \Gamma^{r} \epsilon \bar{\psi_{\mu}} \Gamma_{t1 q
  p n m} \psi_{\rho};
\end{aligned}
\end{cdbout}
\begin{cdbcom}
This is not yet in the desired form, so we join gamma matrices
until we are left with fully anti-symmetrised gamma matrix products, 
\end{cdbcom}
\begin{cdbcont}
converge(obj):
   join_gamma(_)
   distribute(_)
   eliminate_kronecker(_)

canonicalise(_)
rename_dummies(_);
\end{cdbcont}
\begin{cdbout}
\begin{aligned}
& \frac{1}{4}\, T^{\mu \nu \rho} e_{\mu}\,^{m} \bar{\theta} \Gamma^{m n}
\epsilon \bar{\psi_{\nu}} \Gamma_{n} \psi_{\rho} 
+ \frac{5}{16}\, T^{\mu \nu \rho} e_{\mu}\,^{m} \bar{\theta} \epsilon
\bar{\psi_{\nu}} \Gamma_{m} \psi_{\rho} 
+ \frac{3}{32}\, T^{\mu \nu
  \rho} e_{\mu}\,^{m} \bar{\theta} \Gamma^{m n p} \epsilon
\bar{\psi_{\nu}} \Gamma_{n p} \psi_{\rho} \\[1ex]
&+ \frac{1}{4}\, T^{\mu \nu \rho} e_{\mu}\,^{m} \bar{\theta} \Gamma^{n}
\epsilon \bar{\psi_{\nu}} \Gamma_{m n} \psi_{\rho} 
+ \frac{1}{384}\, T^{\mu \nu \rho} e_{\mu}\,^{m} \bar{\theta}
\Gamma^{n p q r} \epsilon \bar{\psi_{\nu}} \Gamma_{m n p q r}
\psi_{\rho};
\end{aligned}
\end{cdbout}
\begin{cdbcom}
This is indeed equivalent to the right-hand side of~\eqref{e:fierzthing}.
\end{cdbcom}
\botrule Note in particular once more the simplicity of the input on
line~12-14. As advertised, very little knowledge of the program is
needed in order to be able to read and follow a calculation in a \Cdb
notebook.

\subsection{Other assorted topics}
\label{s:other} 

So far we have only seen the substitution command in action on bosonic
objects. However, the substitution command in \Cdb is aware of
anti-commuting objects as well, and will take care of signs whenever products
in the pattern and the expression are not ordered in the same way. It
will also refuse to match a pattern if two symbols are declared
non-commuting, and do not appear in the same order in the pattern and
in the expression. 

Anti-commutativity comes in two flavours: self-anticommutativity and
mutual anti-com\-muta\-tivity. The first is used for objects which carry
an index: if $\psi_{\mu}$ is declared self-anticommuting, it means
that $\psi_{\mu}\psi_{\nu} = - \psi_{\nu} \psi_{\mu}$. Below is an
example to illustrate these concepts and the substitution
functionality: \toprule
\begin{cdbin}
\psi_{\mu}::SelfAntiCommuting.
{ \chi, \psi_{\mu} }::AntiCommuting.
ex:= \chi A^{\mu\nu} \psi_{\mu} \chi \psi_{\nu};
\end{cdbin}
\begin{cdbout}
\chi A^{\mu\nu} \psi_{\mu} \chi \psi_{\nu};
\end{cdbout}
\begin{cdbin}
substitute(_, $\psi_{\mu} \psi_{nu} -> B_{\mu\nu}$ );
\end{cdbin}
\begin{cdbout}
- \chi \, A^{\mu \nu} \, B_{\mu \nu} \, \chi;
\end{cdbout}
\begin{cdbcom}
By declaring $\chi$ and $\psi_{\mu}$ to be mutually anti-commuting, 
\Cdb knows that a sign should be picked up when doing the
substitution. 

In order to illustrate the meaning of \verb|SelfAntiCommuting|, we
declare $A^{\mu\nu}$ to be symmetric and canonicalise the
expression,
\end{cdbcom}
\begin{cdbin}
A^{\mu\nu}::Symmetric.
ex:= \chi A^{\mu\nu} \psi_{\mu} \chi \psi_{\nu};  
canonicalise(_);
\end{cdbin}
\begin{cdbout}
0;
\end{cdbout}
\begin{cdbcom}
Because $\psi_{\mu}$ has been declared \verb|SelfAntiCommuting|, the
program knows that $\psi_{\mu}\psi_{\nu}$ is anti-symmetric in its two
indices, and has used this to simplify the expression.
\end{cdbcom}
\botrule

To conclude, let us discuss one more calculation which combines some
of the functionality of \Cdb discussed above, and also shows how to
handle variational problems, especially those involving fermionic
objects. An example is the variation of a Lagrangian under
supersymmetry transformations.  Consider the following sample
calculation,
\eject

\toprule
\begin{cdbin}
def post_process(ex):
   eliminate_kronecker(_)
   sort_product(_)
   collect_terms(_)
\end{cdbin}
\begin{cdbcont}
D{#}::Derivative.
\bar{#}::DiracBar.
\delta{A??}::Derivative.
{m,n,p,q,r,s,t,u,m#}::Indices(flat).
{m,n,p,q,r,s,t,u,m#}::Integer(0..3).
{\mu,\nu,\rho,\sigma,\kappa,\lambda,\alpha,\beta}::Indices(curved,position=fixed).
{\mu,\nu,\rho,\sigma,\kappa,\lambda,\alpha,\beta}::Integer(0..3).
\end{cdbcont}
\begin{cdbcom}
Declaration of bosonic fields,
\end{cdbcom}
\begin{cdbcont}
e^{m \mu}::Vielbein.
e_{m \mu}::InverseVielbein.
g^{\mu\nu}::InverseMetric.
g_{\mu\nu}::Metric.
\end{cdbcont}
\begin{cdbcom}
Declaration of fermionic fields,
\end{cdbcom}
\begin{cdbcont}
{ \epsilon,\psi_{\mu},\psi_{\mu\nu} }::Spinor(dimension=4, type=Majorana).
\Gamma_{#{m}}::GammaMatrix(metric=\delta).
{ \psi_{\mu\nu}, \psi_{\mu}, \epsilon }::AntiCommuting.
{ \psi_{\mu}, \psi_{\mu\nu} }::SelfAntiCommuting.
{ \epsilon, \psi_{\mu}, \psi_{\mu\nu} }::SortOrder.
\Gamma_{#}::Depends(\bar{#}).
\psi_{\mu\nu}::AntiSymmetric.
\end{cdbcont}
\begin{cdbcom}
Input of the Lagrangian and rewriting such that all indices on gamma
matrices are flat,
\end{cdbcom}
\begin{cdbcont}
L:= -1/2 e e^{n \nu} e^{m \mu} R_{\mu\nu n m} 
                 - 1/2 e \bar{\psi_\mu} \Gamma^{\mu\nu\rho} D_{\nu}{\psi_{\rho}}:
rewrite_indices(_, $\Gamma^{m n p}$, $e^{n \mu}$);
\end{cdbcont}
\begin{cdbout}
L := -\frac{1}{2} R_{\mu\nu n m} e e^{m\mu} e^{n \nu} - \frac{1}{2}
 \bar{\psi}_\mu \Gamma^{mnp} D_{\nu}\psi_{\rho} e e^{m \mu} e^{n \nu}
 e^{p \rho}\,;
\end{cdbout}
\begin{cdbcom}
In the 1.5th~order formalism, the supersymmetry transformation rules
are given by
\end{cdbcom}
\begin{cdbcont}
susy:= { e^{n \mu} -> -\bar{\epsilon} \Gamma^m \psi_\nu e^{m \mu} e^{n \nu},
         e         -> e \bar{\epsilon} \Gamma^n \psi_\mu e^{n\mu},
         \psi_\mu  -> D_{\mu}{\epsilon} }:
\end{cdbcont}
\begin{cdbcom}
Varying under supersymmetry is now a simple matter of using the
\verb|vary| command, giving it the rules defined above. This will
automatically assume infinitesimal variations.
\end{cdbcom}
\begin{cdbcont}
vary(L, susy);
\end{cdbcont}
\begin{cdbout}
\begin{aligned}
L := & - \frac{1}{2}\, R_{\mu \nu n m} \overline{\epsilon} \Gamma^{p} \psi_{\rho} e e^{p \rho} e^{m \mu} e^{n \nu} 
+ \frac{1}{2}\, R_{\mu \nu n m} e \overline{\epsilon} \Gamma^{p} \psi_{\rho} e^{p \mu} e^{m \rho} e^{n \nu} \\
&+ \frac{1}{2}\, R_{\mu \nu n m} e e^{m \mu} \overline{\epsilon} \Gamma^{p} \psi_{\rho} e^{p \nu} e^{n \rho} 
 - \frac{1}{2}\, \Gamma^{m n p} \overline{D_{\mu}{\epsilon}}
D_{\nu}{\psi_{\rho}} e e^{m \mu} e^{n \nu} e^{p \rho} \\
& - \frac{1}{2}\, \Gamma^{m n p} \overline{\psi_{\mu}}
D_{\nu}{D_{\rho}{\epsilon}} e e^{m \mu} e^{n \nu} e^{p \rho} 
- \frac{1}{2}\, \Gamma^{m n p} \overline{\psi_{\mu}} D_{\nu}{\psi_{\rho}}
\overline{\epsilon} \Gamma^{q} \psi_{\sigma} e e^{q \sigma} e^{m \mu} e^{n
  \nu} e^{p \rho} \\
&+ \frac{1}{2}\, \Gamma^{m n p} \overline{\psi_{\mu}} D_{\nu}{\psi_{\rho}} e
\overline{\epsilon} \Gamma^{q} \psi_{\sigma} e^{q \mu} e^{m \sigma} e^{n
  \nu} e^{p \rho} 
+ \frac{1}{2}\, \Gamma^{m n p} \overline{\psi_{\mu}} D_{\nu}{\psi_{\rho}} e
e^{m \mu} \overline{\epsilon} \Gamma^{q} \psi_{\sigma} e^{q \nu} e^{n
  \sigma} e^{p \rho} \\
&+ \frac{1}{2}\, \Gamma^{m n p} \overline{\psi_{\mu}} D_{\nu}{\psi_{\rho}} e e^{m \mu} e^{n \nu} \overline{\epsilon} \Gamma^{q} \psi_{\sigma} e^{q \rho} e^{p \sigma};
\end{aligned}
\end{cdbout}
\botrule 
(The rest of the calculation, in which the result is rewritten using
partial integration and a Fierz identity to show that it vanishes, can
be found in a more extensive notebook on $N=1$ supergravity in four
dimensions, available from the web site).

Various aspects of \Cdb are visible here: the use of multiple
index types and conversion between them, the user-specified sort order
of fields which takes into account commutativity properties of
tensors, the pattern matching routines which automatically deal with dummy
index relabelling, the readability of input/output and so on.

\section{Conclusions}

We have discussed the capabilities of the new computer algebra system
``\Cdb'', by applying it to a number of concrete calculations. There are
several aspects of this system which make it particularly well-suited
to solve field-theory problems. Firstly, it uses (a subset of)~\TeX{}
not only for output, but also for input. Compared to other computer
algebra systems, this makes it much easier to input complicated
expressions without errors, as there is hardly any new syntax to be
learnt. Secondly, the program has built-in facilities for many
concepts which occur in field theory, like anti-commuting variables,
gamma matrix algebra, implicit dependence on coordinates, accents,
multiple dummy index sets, canonicalisation of tensor expressions and
so on. Although there exist other systems which share some of the
functionality of \Cdb, the implementation in \Cdb was designed from
scratch so as to make problem solving resemble as close as possible
the steps one would follow with pencil and paper.

The program is entirely built on freely distributable software
libraries, i.e.~it does not make use of any proprietary computer
algebra system.  The system is available in source code for all
supported platforms, as well as in binary form for Linux and Windows
computers; the reader is referred to the web
site~\url{http://cadabra.science} for download and installation
instructions as well as the help forum.

Given the program's scope and size, there of course remains quite some
room for improvements and additions. An upcoming paper will describe
the improvements introduced in the 2.x series.
\vfill\eject

\section*{Acknowledgements}

The development of Cadabra was made possible by support from
DAMTP at Cambridge University, from CERN and in particular from the
Albert-Einstein-Institute in Potsdam.  This work was also sponsored in
part by VIDI grant 016.069.313 from the Dutch Organisation for
Scientific Research (NWO).  I am grateful to the Department of
Mathematical Sciences at Durham University for hospitality while this
work was being completed. Many thanks to Marcus Berg for comments on
the first version of this paper.

\begingroup\raggedright\endgroup
\end{document}